# Experimental observation of sub-radiant longitudinal mode and anomalous optical transmission in metallic wire meshes

Weijie Dong[1], Xiaoxi Zhou[1], Xinyang Pan[1], Haitao Li[1], Gang Wang[1], Yadong Xu[1], and Bo Hou[2]

[1] *School of Physical Science and Technology & Collaborative Innovation Center of Suzhou Nano Science and Technology, Soochow University, 1 Shizi Street, Suzhou 215006, China*

[2] *Wave Functional Metamaterial Research Facility, The Hong Kong University of Science and Technology (Guangzhou), 1 Duxue Road, Guangzhou 511400, China*

**Abstract**

In conventional plasmonic media and plasmonic metamaterials, such as metallic wire mesh, longitudinal mode of electromagnetic wave manifests itself in frequency overlapping transverse modes, which impedes clear observation of longitudinal-mode-specific physical effects. Through interlacing two sets of wire meshes, an ideal band for longitudinal mode is achieved ranging from zero frequency to plasma frequency where transverse modes are completely forbidden. The unique spectral separation of modes facilitates the observation of pure longitudinal mode and related plasmonic effects in bulk medium. We report the microwave experiment of anomalous optical transmission, induced solely by electromagnetic longitudinal mode resonance, below plasma frequency in such wire mesh medium.

Authors to whom correspondence should be addressed: [Yadong Xu, ydxu@suda.edu.cn; Bo Hou, bohou@hkust-gz.edu.cn]

In the study of light propagation through artificially designed periodic structures, two anomalous transmission phenomena have attracted intense interests and longstanding investigation [Refs. 1-15]. One of them is associated with subwavelength hole arrays [Ref. 1] and the other is attributed to negative index of refraction [Ref. 2], giving birth to two research fields, plasmonics and metamaterials [Refs. 16-22]. Seemingly extraordinary transmission in a composite medium of metallic wires [Ref. 23] and split-rings [Ref. 24] is understood being the transparency of metamaterial with simultaneously negative permittivity and permeability [Ref. 2]. It is well known that single negative material presents a stopband for light or electromagnetic (EM) wave whenever its permittivity $\varepsilon$ or permeability $\mu$ is negative. However, double negative material that combines $\varepsilon < 0$ and $\mu < 0$ material brings about a passband, therefore giving rise to the light transparency which defines negative index of refraction ($n < 0$), as illustrated in Fig. 1(a).

One conventional example of $\varepsilon < 0$ metamaterial is the metallic wire mesh where the mesh size defines essentially the plasma frequency $\omega_p$. The optical transparency is prohibited below the plasma frequency, and therefore the passband is not expected when combining such two metallic wire meshes, which seems like the composition of two $\varepsilon < 0$ material. Remarkably, theoretical calculation has found the anomalous transmission of light through two sets of metallic wire meshes when them are composited in an interlaced way [Ref. 25]. Being different from the single set of metallic wire mesh, the interlaced metallic meshes embody a low frequency longitudinal mode (LM) down to zero frequency (DC) and realize an ultra-broad passband effect below a modified plasma frequency $\Omega_p$ (blue-shifted in spectrum with respect to $\omega_p$) [Refs. 26-28], as depicted in Fig. 1(b).

Interestingly, the LM appears in the ultra-broad spectral range that is exclusive to the transverse modes (TMs), and therefore the unique spectral feature is ideal to investigate the pure LM physics. Given the LM band and its negative dispersion, two straightforward wave effects can be expected in the bulk metamaterial, one being optical transmission [Ref. 25] and the other being negative refraction [Ref. 27]. In the letter, we fabricate the kind of microwave metamaterial which is composed of two interlaced metallic meshes with each mesh being simple cubic lattice architecture. Inside the composite medium, an ideal LM band is achieved, where only LM takes place without transverse modes overlapping. Consequently, anomalous optical transmission through the metamaterial slab in free space is observed for p-polarized wave and is attributed to the Fabry-Perot (FP) mechanism of sub-radiant LM. Although the interlaced metallic wire meshes have been made recently via top-down three-dimensional printing technology, the extrinsic antenna-like stubs must be designed on the metamaterial surfaces to couple the LM with the incident wave in free space [Ref. 29]. In fact, the LM excitation and the transmission observation can be intrinsically done via oblique incidence without any addition of extra surface stubs to the metamaterial, as shown here.

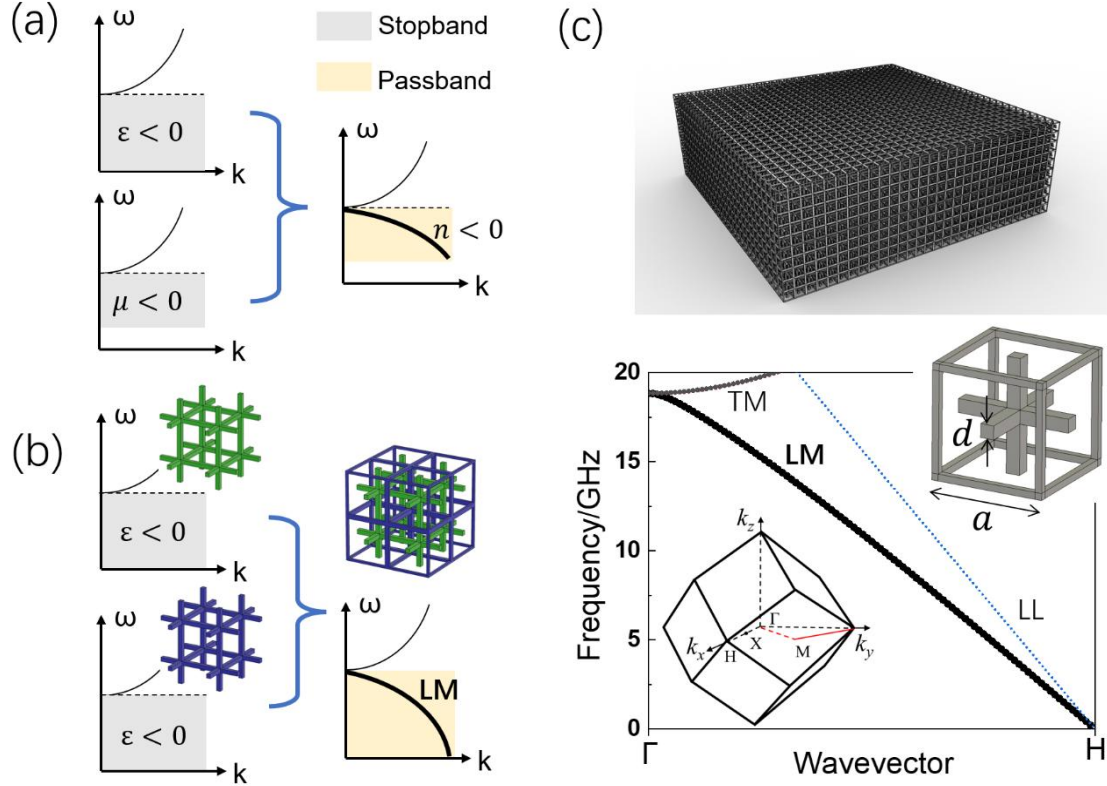

Fig. 1: (a) Sketch of transparency due to negative refraction, where thin solid lines and dash lines denote the transverse modes and the (local) longitudinal modes, respectively. (b) Sketch of transparency in the interlaced metallic wire meshes with identical component mesh (green one and purple one), where a longitudinal mode dispersion appears below the modified plasma frequency $\Omega_p$, and thin solid lines and dash lines denote the transverse modes and the (local) longitudinal modes, respectively. (c) A schematic model, where a lattice cell is illustrated with simple cubic periodicity $a$ and wire size $d$, and two meshes are identical. The band diagram is plotted along <100> direction, and the four points labelled in the Brillouin zone (inset) are Γ at $(0,0,0)$, H at $\frac{2\pi}{a}(1,0,0)$, X at $\frac{\pi}{a}(1,0,0)$, and M at $\frac{\pi}{a}(1,1,0)$. The light line (LL) is shifted deliberately towards H to facilitate the comparison of the slope between the longitudinal mode and the light line.

First, we present a numerical model to illustrate the aforementioned LM band. As shown in Fig. 1(c), two metallic meshes are interlaced to form a bulk of structure. The two meshes are identical to each other and both are in simple cubic lattice with periodicity $a = 10mm$ for the lattice cell and size $d = 2mm$ for the square cross-section of wire. In the case of being identical meshes and interlacing along <111> direction by $(\frac{a}{2},\frac{a}{2},\frac{a}{2})$, the composing structure is in body-centered cubic lattice. The calculated dispersion along <100> direction displays a nondegenerate band covering from DC to plasma frequency $\frac{\Omega_p}{2\pi} \approx 18GHz$. In calculation, metal is approximated as perfect electric conductor (PEC). Furthermore, the band is a cone centered at the corner (H) of the Brillouin zone (BZ), and the cone has been justified as the electromagnetic LM [Ref. 28]. The existence of the LM will give rise to the transmission of EM wave in otherwise forbidden frequency range. However, it should be emphasized and distinguished that the longitudinal wave is mediating the physical effect, in sharp contrast to the transverse wave governing in the $n < 0$ passband, as compared in Figs. 1(a) and 1(b).

Next, we use the layer-by-layer assembling strategy to fabricate the interlaced wire mesh sample. An aluminum plate is machined into a layer structure that is a two-dimensional periodic array of square holes with in-plane periodicity $a_x = a_y = a = 10mm$, in-plane size of holes $a_x - d_x = a_y - d_y = a - d = 8mm$ and out-of-plane depth of holes $d = 2mm$, and additionally a square rod with out-of-plane length $(a - d)/2$ is machined periodically extruding on each side of the layer structure, as illustrated in **Supplementary Material (Section A)**. Then, we cascade such structure in layer-by-layer way with every one layer offset along the in-plane <11> direction. In the end, the cable ties are applied to fasten the whole assembled structure. The fabricated sample measures $\sim 50cm \times 50cm \times 25mm$ in overall size and shows a flatten front/back surface without rod extrusion on both [001] faces, as shown in Fig. 2(a). In experiment, the microwave transmission is measured in free space at frequencies from 1 to 18GHz with a network analyzer and two broadband horn antennas. The p- and s-polarized wave is incident at the oblique angle $\theta^{inc}$, respectively, and the sample is rotated about the z-axis, i.e. <001> direction, by the azimuthal angle $\varphi$ for different incident planes, as depicted in Fig. 2(a).

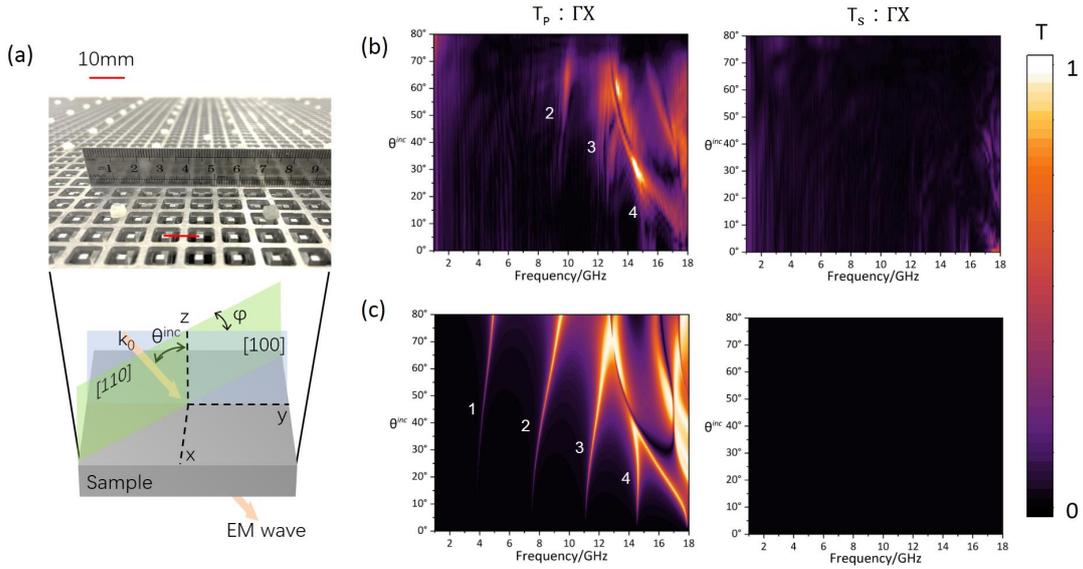

Fig. 2: (a) Upper: photo of the interlaced structure of two identical wire meshes, each being the simple cubic lattice with lattice constant $a = 10mm$ and the wire size $d = 2mm$; Lower: sketch of transmission measurement where the plane electromagnetic wave with free space wavenumber $k_0 = \omega/c$ ($c$ being the speed of light in vacuum) is impinging at the incident angle $\theta^{inc}$, and the incident plane is rotated about the z-axis, i.e., <001> direction, by an angle $\varphi$ to characterize different planes, such as [100] and [110]. (b) Experimental transmission results, respectively, under the incidence of p- and s-polarized wave for ΓX direction. (c) The corresponding simulated results.

The polarization-resolved transmission, $T_p(\omega, \theta^{inc})$ and $T_s(\omega, \theta^{inc})$, as a function of wave
6

frequency and incident angle is firstly measured when EM wave is incident within the [100] incident plane, and the results are plotted in Fig. 2(b), respectively, for two polarizations. Due to the plasma frequency $\frac{\Omega_p}{2\pi} \approx 18\text{GHz}$, transmission is almost zero for s-polarization below ~18GHz. In contrast, significant transmission and several peaks, as labeled, are observed in the case of p-polarization, because of the excited plasmon mode response. The rapid reduction of transmission beyond $\theta^{inc} \sim 70$ degree is due to the undersize effect of the sample with large incident angle. It is noted that the transmission peaks become narrow and disappear when the incident angle decreases. The experimental results are verified by the simulation, seeing Fig. 2(c). An extra sharp transmission peak that fails to be resolved in experiment due probably to the assembling precision is observed at the lower frequencies in simulation. And the four peaks, 1, 2, 3, and 4, fade with decreasing the angle, which implies the sub-radiant property of the plasmon mode under the normal incidence, and therefore these plasmon mode responses become hardly excited at small $\theta^{inc}$ in despite of p-polarization. This is the LM evidence in observation because longitudinal polarization leads to the complete decoupling of LM with the incident transverse wave at normal incidence. Additionally, the peaks exhibit the approximately constant separation in frequency, which is a spectral feature of FP resonance.

The LM cone in the BZ is obtained by measuring the transmission along the different incident plane characterized by the azimuthal angle φ, which denotes a rotation about the z-axis. In Fig. 3, the measured and simulated results are plotted for ΓM direction, i.e., φ=45 degree, as indicated by the [110] plane in Fig. 2(a). Although more transmission noise appears at the low frequency region, the labeled peaks are still identified and appear only for p-polarization. The transmission features persist to other directions, φ=15 and 30 degrees, which is verified via measurement and simulation in **Supplementary Material (Section B)**. Moreover, the LM and the transmission features are also robust to lossy metal with electrical conductivity $\sigma$ ranging from good conductor ($\sigma \sim 10^7$ S/m) to moderate conductor ($\sigma \sim 10^5$ S/m), seeing **Supplementary Material (Section C)**.

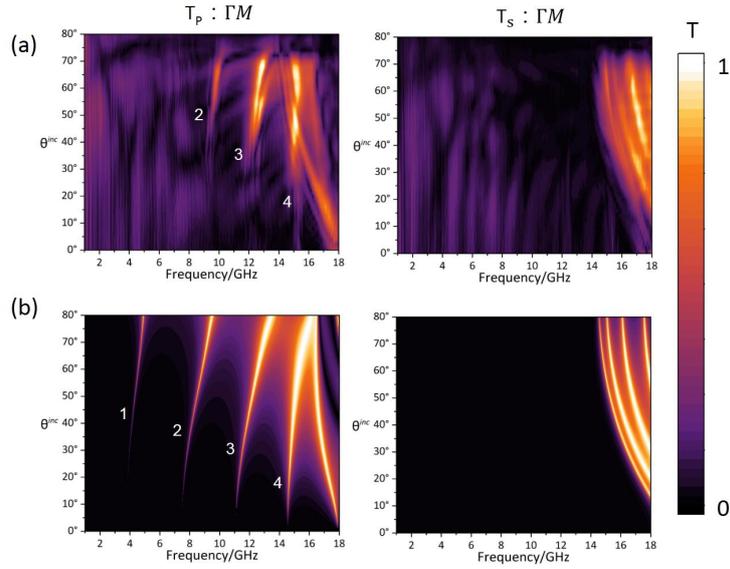

Fig. 3: (a) Experimental transmission results, respectively, under the incidence of p- and s-polarized wave for ΓM direction, as shown by the [110] plane in Fig. 2(a). (b) The corresponding simulated results.

The EM transmission is attributed to FP resonance of LM in the interlaced sample below the plasma frequency where any transverse mode is forbidden, seeing Ref. [25] in which the authors emphasize that the two component meshes, labeled as A and B, must be asymmetric to enable the anomalous transmission. Notably, for the sample fabrication, two sets of meshes are modeled and designed to be identical and be symmetrically located, and nevertheless the anomalous transmission is still observed. In fact, for the practical sample, both components are no longer symmetric in exact sense, due to various error in machining, positioning and assembling of the interlaced structure. Consequently, the sub-mesh symmetry is broken, which causes a band folding and gap [Refs. 30-31], as shown in **Supplementary Material (Section D)**. The LM is folded from the corner H to the center Γ in the BZ and forms two branches that are the low-frequency one with positive dispersion and the high-frequency one with negative dispersion. The *n*th-order FP transmission frequencies $\omega_n$ can be analyzed through an effective medium model [Ref. 25, also seeing **Supplementary Material (Section E)**]:

$$\frac{\omega_n^2}{c^2/l_0} = \left(\frac{\omega_n}{c}\sin\theta^{inc}\right)^2 + \left(\frac{n\pi}{L}\right)^2,$$

where $l_0$ is the slow wave factor in the nonlocal plasma model, $L$ is the thickness of the sample, and the integer $n = 1, 2, 3 \cdots$. In addition, the longitudinal polarization nature of LM has determined that the FP resonance cannot be excited by the normally incident plane wave even though the wave is p-polarized, which is a consequence of orthogonal orientation of electric fields and therefore gives rise to the fading peaks with the decreasing incident angle.

In conclusion, the interlaced wire mesh structure displays an ideal band of LM from DC to plasma frequency where the transverse wave is absent. The transmission induced merely by the LM resonance is therefore observed in free space, and the sub-radiant state of LM manifests under oblique incidence. The interlaced metamaterial will provide to a platform to explore LM-specific electromagnetic phenomena, e.g., negative refraction of longitudinal EM wave. In particular, the ultra-broad spectrum of the ideal LM band can be utilized to demonstrate broadband all-angle negative refraction. Additionally, the characteristic of single mode with large wavevector in the case of sub-mesh symmetry will be beneficial to design subwavelength devices like resonator and filter for technology applications.

**Supplementary Material**

The supplementary material includes the photos of the sample and the setup, the effect of different incident planes, the effect of different conductivities, the sub-mesh symmetry breaking, and the Fabry-Perot resonance of the longitudinal mode.

**Acknowledgments**

This work was supported by the Natural Science Foundation of China (NSFC) (Grant No. 12074279), the Major Program of Natural Science Research of Jiangsu Higher Education Institutions (Grant No.18KJA140003), and the Priority Academic Program Development (PAPD) of Jiangsu Higher Education Institutions.

Supplementary Material for

# Experimental observation of sub-radiant longitudinal mode and anomalous optical transmission in metallic wire meshes

Weijie Dong[1], Xiaoxi Zhou[1], Xinyang Pan[1], Haitao Li[1], Gang Wang[1], Yadong Xu[1], and Bo Hou[2]

[1] *School of Physical Science and Technology & Collaborative Innovation Center of Suzhou Nano Science and Technology, Soochow University, 1 Shizi Street, Suzhou 215006, China*
[2] *Wave Functional Metamaterial Research Facility, The Hong Kong University of Science and Technology (Guangzhou), 1 Duxue Road, Guangzhou 511400, China*

**A. The photos of the sample and the setup**

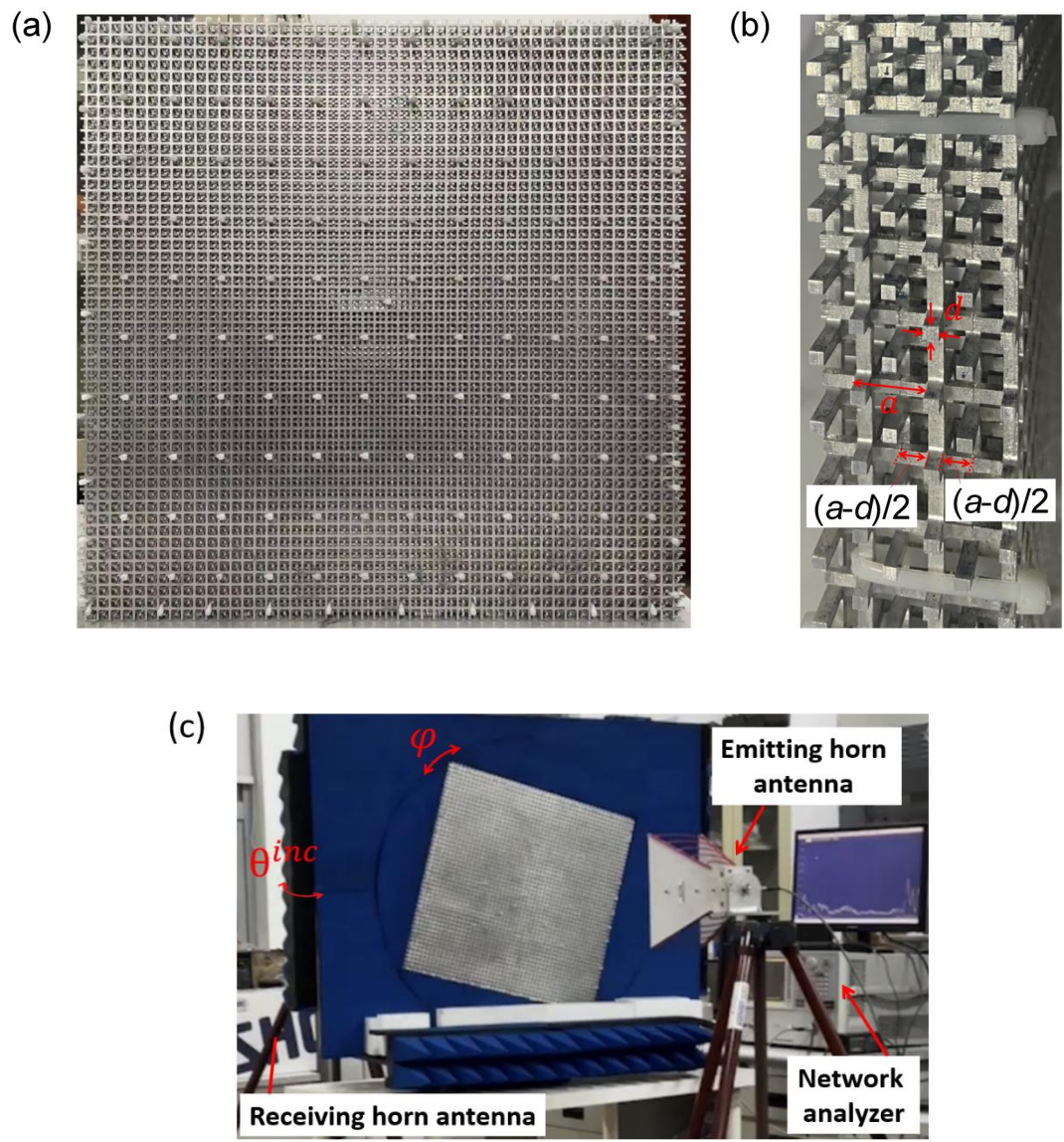

Fig. S1: (a) Front view and (b) zoomed side view of the sample that measures $\sim 50cm \times 50cm$ in transvers size and $\sim 25mm$ in thickness. The simple cubic lattice cell has the periodicity $a = 10mm$, and the wire has the square cross-section with the size $d = 2mm$. (c) The experimental setup where the receiving horn antenna is blocked and only its tripod is seen.

## B. The effect of different incident planes

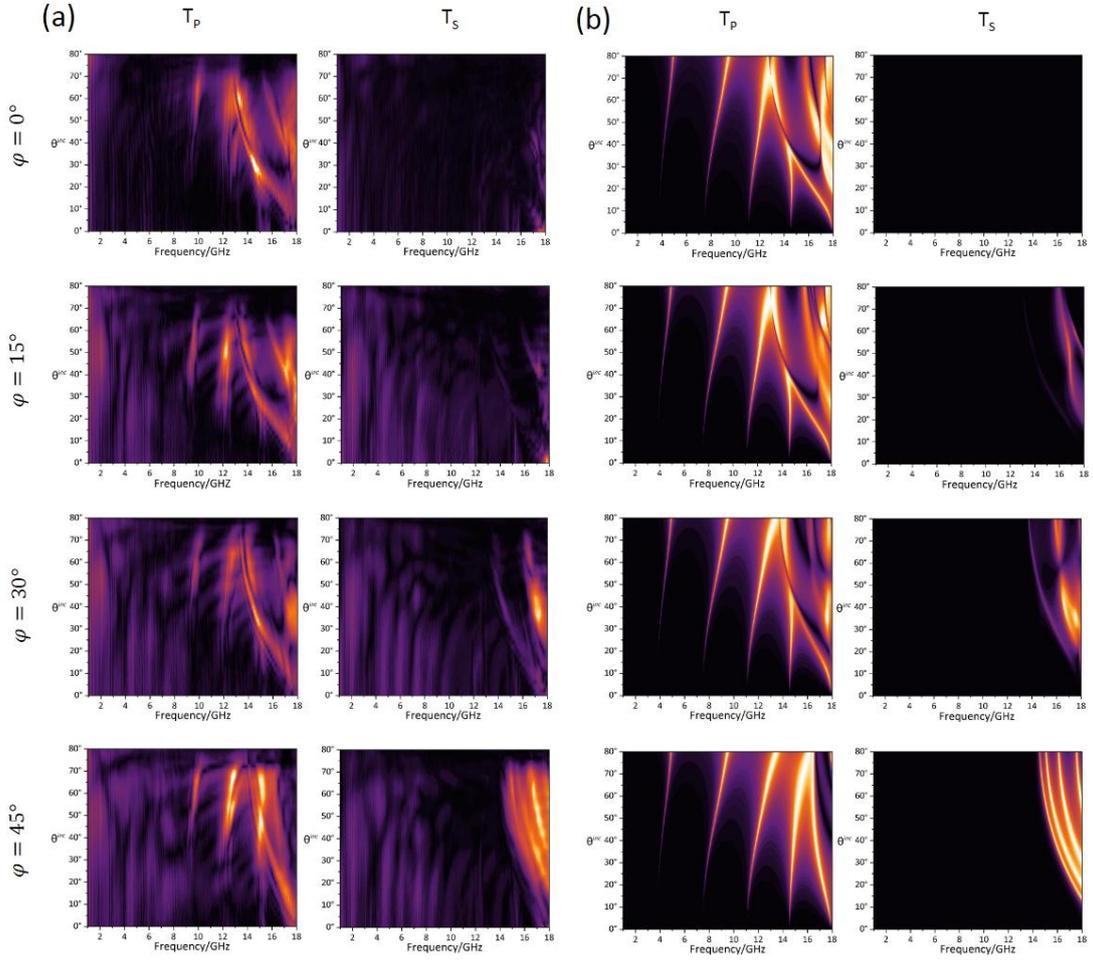

Fig. S2: (a) Measured and (b) simulated transmission, $T_p(\omega, \theta^{inc})$ and $T_s(\omega, \theta^{inc})$, of the sample at the different incident planes characterized by the azimuthal angle $\varphi$.

## C. The effect of different conductivities

Our microwave simulations show good conductors like copper, aluminium, and steel, the order of magnitudes of whose conductivities $\sigma$ are greater than $10^5$ S/m, can be approximated by PEC. The simulated transmissions in the case of good conductors are almost the same as the PEC case. Moderate and poor conductors with the order of magnitudes of conductivities being $10^5$ S/m (e.g., nichrome), $10^4$ S/m (e.g., conductive composite), and $10^3$ S/m (e.g., amorphous carbon) are also examined and the simulated results in these cases are plotted in Fig. S3. Therefore, the LM and the corresponding transmission effect are robust to a large range of lossy metal from $\sigma \sim 10^7$ S/m to $\sigma \sim 10^4$ S/m.

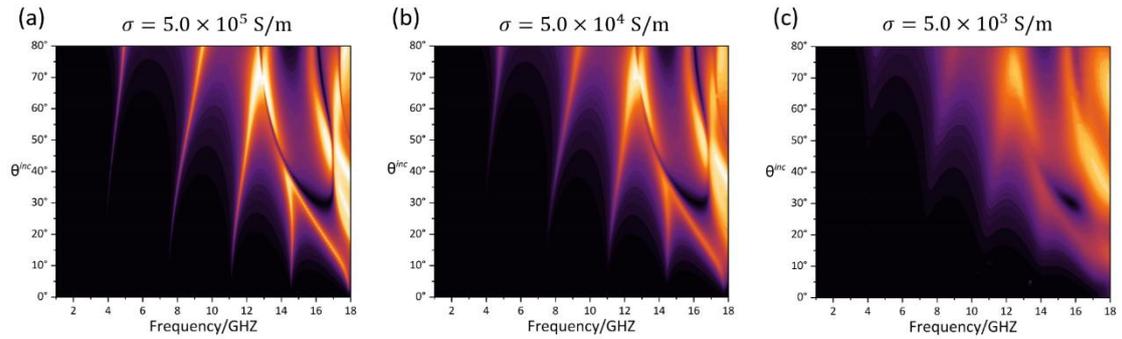

Fig. S3: Simulated transmissions $T_p(\omega, \theta^{inc})$ along ΓX direction ($\varphi = 0°$) for the metamaterial made of lossy conductor with different conductivity $\sigma$ (a) $5.0 \times 10^5$ S/m, (b) $5.0 \times 10^4$ S/m,

and (c) $5.0 \times 10^3$ S/m.

**D.  The sub-mesh symmetry breaking**

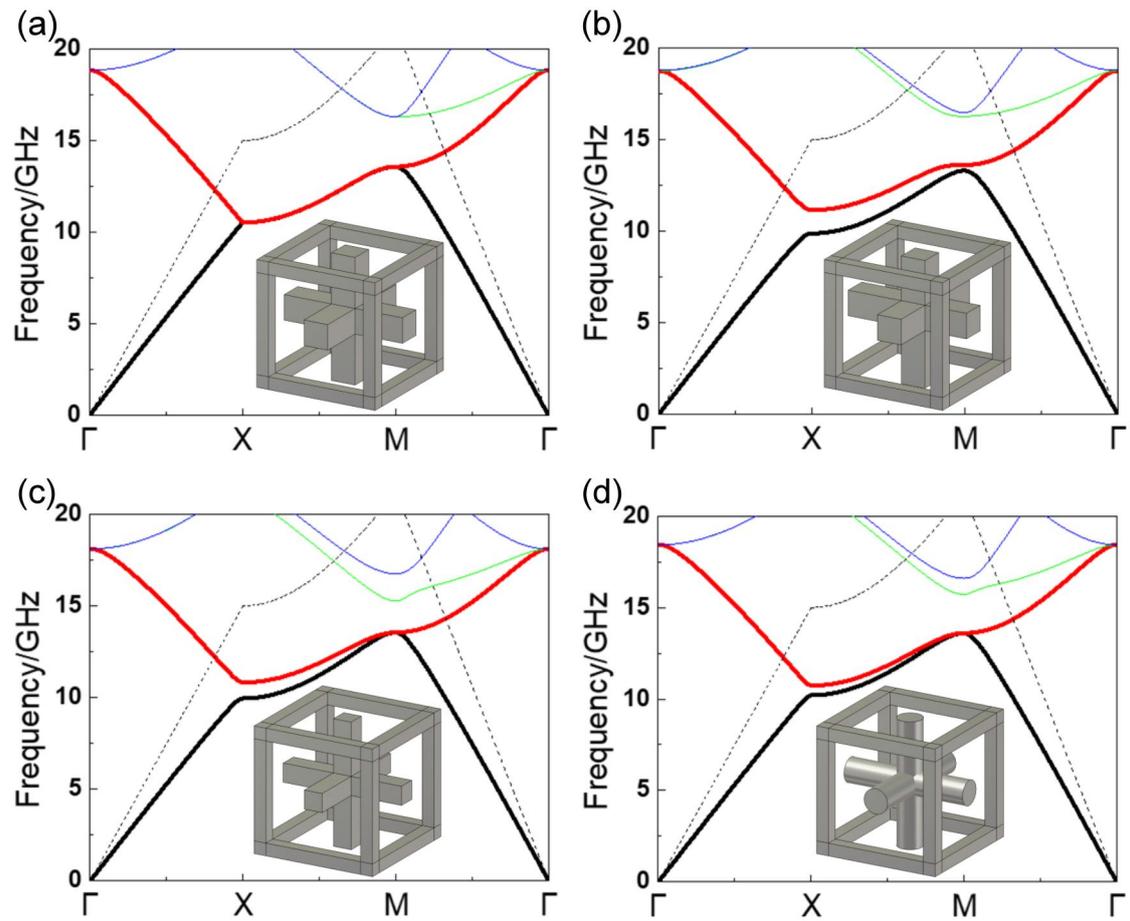

Fig. S4: Band diagrams along high-symmetry path for (a) identical meshes with symmetrical location (center-to-center offset = $(\frac{a}{2}, \frac{a}{2}, \frac{a}{2})$ ), (b) identical meshes with asymmetrical location

(center-to-center offset $\neq (\frac{a}{2},\frac{a}{2},\frac{a}{2})$), (c) different meshes (the wire being shrunk in a mesh) with symmetrical location, and (d) different meshes (the wire being rounded in a mesh) with symmetrical location. The thick black and thick red lines denote the low-frequency and high-frequency LM, respectively, the dash line is the light line in vacuum, and the inset shows the lattice cell in simulation.

### E. The Fabry-Perot resonance of the longitudinal mode

For the interlaced structure, the effective permittivity of the LM within PEC approximation can be expressed [Ref. 25 in main text]:

$$\varepsilon_l(\omega,k) = 1 - \frac{\left(\omega_p^A\right)^2}{\omega^2 - \frac{c^2}{l_0^A}k^2} - \frac{\left(\omega_p^B\right)^2}{\omega^2 - \frac{c^2}{l_0^B}k^2}$$

where $\omega_p^A$ and $\omega_p^B$ are the plasma frequency of the mesh component A and B, respectively, and $l_0^A$ and $l_0^B$ are the dimensionless slow wave factor in the nonlocal plasma model. Solve the condition $\varepsilon_l(\omega,k) = 0$, and we can obtain two solutions ($\pm$) for the LM dispersion:

$$\omega^2 = \frac{1}{2}\left[\frac{c^2}{l_0^A}k^2 + \frac{c^2}{l_0^B}k^2 + \left(\omega_p^A\right)^2 + \left(\omega_p^B\right)^2\right] \pm \frac{1}{2}\sqrt{\frac{c^2}{l_0^A}k^2 - \frac{c^2}{l_0^B}k^2 + \left(\omega_p^A\right)^2 - \left(\omega_p^B\right)^2 + 4\left(\omega_p^A\right)^2\left(\omega_p^B\right)^2}.$$

Here, the ($-$) solution gives us the dispersion of the low-frequency LM and can be simplified, in the case of nearly symmetrical sub-mesh where $\omega_p^A \approx \omega_p^B = \omega_p$ and $l_0^A \approx l_0^B = l_0$, as:

$$\omega \approx \frac{c}{\sqrt{l_0}} k^{(L1)}$$

where $k^{(L1)}$ denotes the wavenumber of the low-frequency LM, and the coefficient $\frac{c}{\sqrt{l_0}}$ plays the role of velocity which is scaled by the slow wave factor with respect to the speed of light $c$.

In our anomalous transmission problem, the dispersion of the low-frequency LM becomes:

$$\frac{\omega^2}{c^2/l_0} = \left(k^{(L1)}\right)^2 = \left(k_y^{(L1)}\right)^2 + \left(k_z^{(L1)}\right)^2$$

with $k_y^{(L1)} = k_0 \sin\theta^{inc} = \frac{\omega}{c}\sin\theta^{inc}$ enforced by the incident wave. Through imposing the Fabry-Perot resonance condition $k_z^{(L1)}L = n\pi$ ($n$ being positive integer and $L=2.5a$ being the thickness of the structure), we obtain the transmission resonance frequency $2\pi f_n$ as a function of $\theta^{inc}$:

$$\frac{(2\pi f_n)^2}{c^2/l_0} = \left(\frac{2\pi f_n}{c}\sin\theta^{inc}\right)^2 + \left(\frac{n\pi}{L}\right)^2.$$

The first four resonances are calculated and plotted in Fig. S5, where $f_n$ appears at 3.8GHz, 7.6GHz, 11.4GHz, and 15.2GHz in the case of $\theta^{inc} = 0$. The comparison with the simulation is illustrated in Fig. S6, and the good agreement between the numerical results and the FP formular is noted, especially for the first two resonances where no high-frequency modes interfere and the effective parameter description is applicable, and displays the FP nature of the transmission peaks.

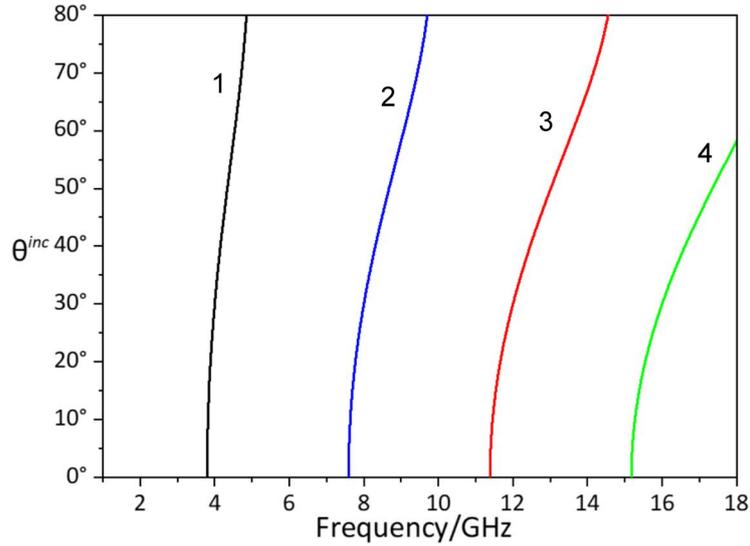

Fig. S5: The relation of the FP resonance frequency $f_n$ versus $\theta^{inc}$, where $n$=1, 2, 3, and 4 as labeled and the slow wave factor $l_0 = 2.5$ is assumed (the value being approximate to the slope ratio between the LM and the light line in Fig. S4).

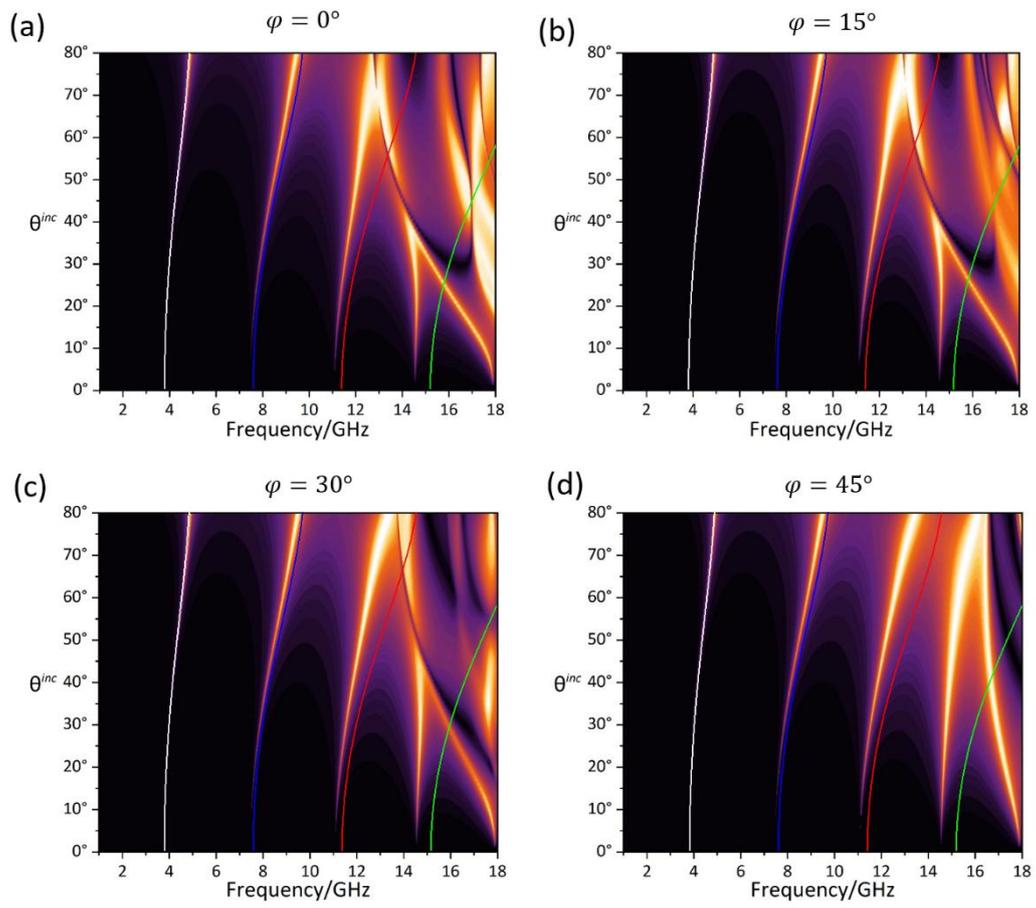

Fig. S6: Comparison between the FP resonance frequency (solid lines) and the simulation results (color maps) for four incident planes with the different angle $\varphi$.